\documentclass[%
 reprint,
 amsmath,amssymb,
 aps,
]{revtex4-2}

\usepackage{graphicx}
\usepackage{dcolumn}
\usepackage{bm}
\usepackage{hyperref}

\usepackage{color}

\begin{document}

\newcommand{\LHL}[1]{\textcolor{blue}{\bf[NOTE: LHL -- #1]}}

\preprint{}

\title{A First Account of the Impact of Ion Electromagnetic Dissociation on Event Exclusivity in Ultraperipheral LHC Collisions}

\author{M. Dyndal}
\email{Mateusz.Dyndal@cern.ch}
 \affiliation{AGH University of Krakow, Faculty of Physics and Applied Computer Science, al. Mickiewicza 30, 30-059 Krakow, PL}
\author{L. A. Harland-Lang}%
 \email{l.harland-lang@ucl.ac.uk}
\affiliation{Department of Physics and Astronomy, University College London, London, WC1E 6BT, UK
}%

\date{\today}

\begin{abstract}
In this Letter we explore the modelling of hadron production in electromagnetic ion dissociation (EMD) processes in high-energy ultraperipheral collisions at LHC energies. Since  EMD  can accompany exclusive particle production in these interactions, we demonstrate that the resulting hadrons can break the exclusivity vetos typically imposed by experiments. 
As two representative examples, we calculate the impact on existing LHC measurements of exclusive muon pair production ($\gamma\gamma\to\mu\mu$) and exclusive coherent $J/\psi$ production. We demonstrate that accounting for this effect resolves long-standing tensions between theoretical predictions and experimental measurements.

\end{abstract}

\maketitle


\textit{Introduction---}In ultrarelativistic heavy-ion collisions at the Large Hadron Collider (LHC), the ion beams have large electric charges $Z$ and are therefore accompanied by strong electromagnetic (EM) fields. 
At impact parameters $b$ larger than twice the nuclear radius, EM-induced reactions become the dominant interaction mechanism. These so-called ultraperipheral collisions (UPCs) provide a clean environment to study various photon--nucleus ($\gamma A$) and photon--photon ($\gamma\gamma$) interactions~\cite{Baltz:2007kq,Klein:2020fmr}. 

A particularly clean signature of UPCs is exclusive production, where the final state of interest is produced with no additional particle activity in the detector; a topical example is the production of lepton pairs, $\gamma\gamma \to \ell\ell~(\ell=e,~\mu,~\tau)$, which has been measured by the ATLAS~\cite{ATLAS:2020epq, ATLAS:2022srr, ATLAS:2022ryk}, CMS~\cite{CMS:2020skx,CMS:2022arf, CMS:2024bnt}, and ALICE~\cite{ALICE:2013wjo, ALICE:2023mfc} collaborations.
Other exclusive photon--initiated ($\gamma\gamma$) processes accessible at the LHC include very rare light-by-light scattering~\cite{ATLAS:2020hii,CMS:2024bnt}.
In the case of exclusive photonuclear ($\gamma A$) interactions, there are a wealth of LHC measurements  of the coherent production of  vector mesons, in particular the $\rho^0$~\cite{ALICE:2020ugp,ALICE:2021jnv,LHCb:2025fzk} and $J/\psi$ mesons~\cite{ALICE:2013wjo,ALICE:2012yye,ALICE:2019tqa,ALICE:2021gpt,LHCb:2022ahs,CMS:2023snh,ALICE:2023jgu,ATLAS:2025aav}.
More recently, the coherent production of $\phi$ and $\Upsilon(1S)$ mesons in UPCs has also been measured~\cite{CMS:2025lsm,CMS:2026soy}.

Due to the enormous EM fields present in  UPCs, additional photon exchanges between the colliding
ions are possible, in addition to the production of the process of interest.
As a result, additional particles can be produced~\cite{ATLAS:2025nac,Harland-Lang:2025bkk}, or one or both ions can undergo EM dissociation (EMD)~\cite{Baltz:2009jk}.
The EMD process can be described as resulting from one ion inducing nuclear excitation via $\gamma A \to A^*$ in the other. The excited ion subsequently de--excites, emitting neutrons and other particles in the process. The dominant mode is the giant dipole resonance (GDR)~\cite{Berman:1975tt}, the lowest-energy excitation, which typically results in single-neutron emission. Higher excitations are also possible, leading to the emission of multiple neutrons, possibly accompanied by other particles. In the following, we distinguish between single-ion ($0nXn$) and mutual ($XnXn$) EMD, where $Xn$ refers to the presence of beam-energy neutrons.

As those extra EMD processes typically produce very forward, beam-energy neutrons, they are kept as part of the exclusive signal, as they pass the exclusivity requirements (exclusivity veto) used by the experiments.
However, when the exchanged photon energy in EMD becomes high, the EMD effectively becomes a deeply inelastic $\gamma$--ion interaction, with additional hadrons being produced~\cite{Pshenichnov:1999hw}.

In this Letter, we investigate the modelling of hadron production in the EMD process with high-energy photons, accompanying exclusively produced final states in LHC UPCs.
We will show that those hadrons can break the exclusivity vetos and hence  introduce bias when comparing experimental results (where these effects are often neglected) with theory predictions (where these effects are not explicitly simulated).
We address this by calculating the relevant corrections for selected UPC measurements from LHC experiments, in particular for exclusive muon-pair production ($\gamma\gamma\to\mu\mu$) and exclusive coherent $J/\psi$ production. 


\begin{figure*}
\includegraphics[scale=0.4]{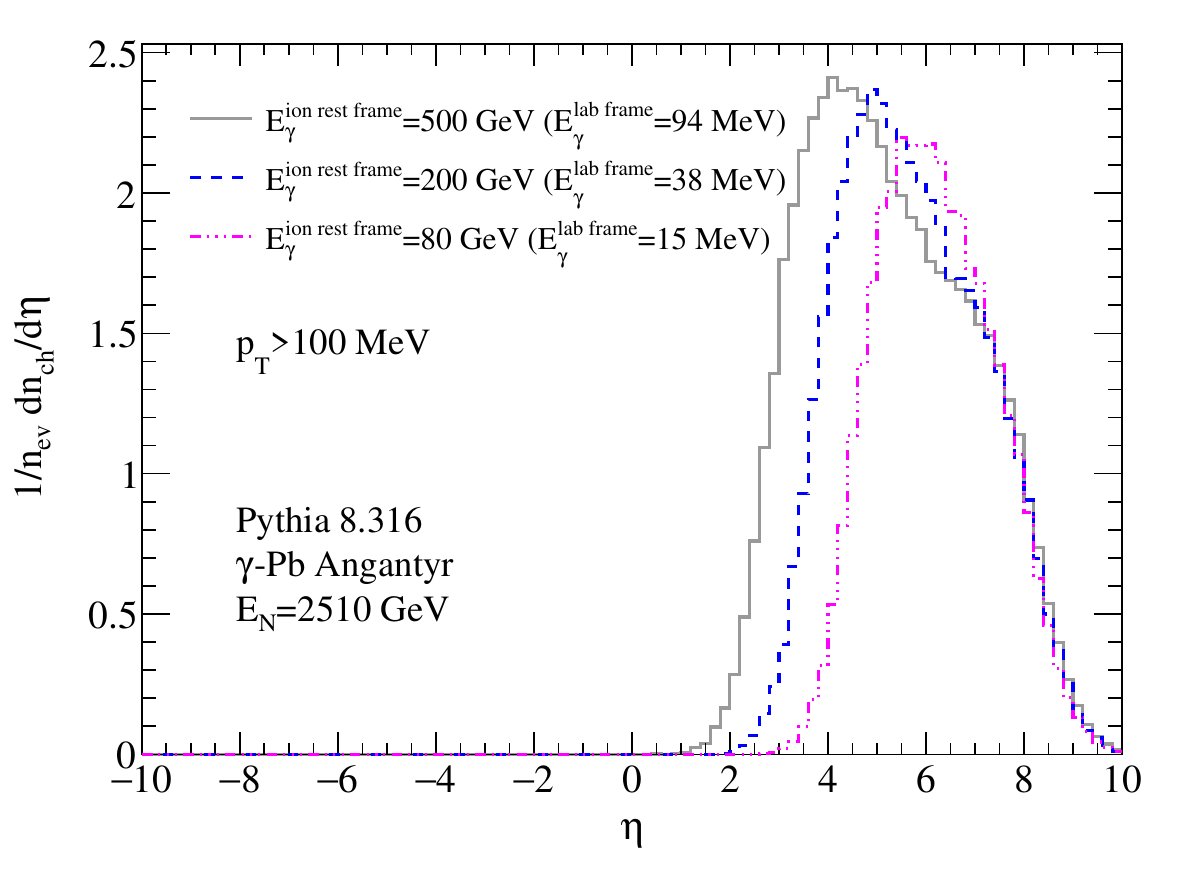}
\put(-120,-5){\textbf{(a)}}  
\includegraphics[scale=0.4]{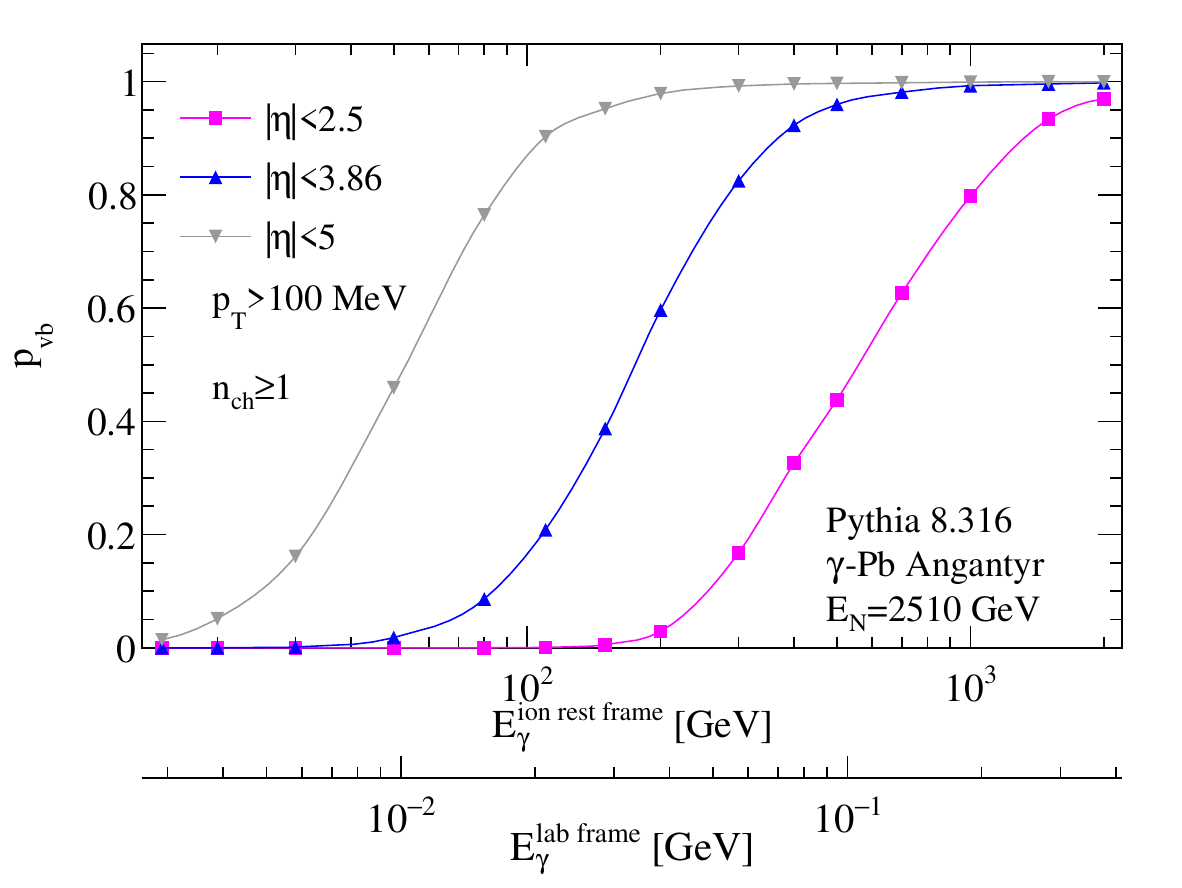}
\put(-120,-5){\textbf{(b)}}  
\caption{\label{p8:dndeta} (a) Charged-particle pseudorapidity distribution from simulated $\gamma Pb$ (EMD) interactions using the \texttt{Pythia8+Angantyr} model. The LHC $Pb$ beam energy of 2510 GeV per nucleon was used, together with several representative photon energies (varied between 15 MeV and 94 MeV). Positive and negative $\eta$ denote the nucleus and photon directions, respectively.  (b) Veto-breaking probabilities for EMD as a function of incident photon energy, calculated in the dissociated ion rest frame. The alternative $x$-axis also shows the photon energy in the laboratory frame. For the veto definition, charged particles with $p_T>100$~MeV and three different values of pseudorapidity requirements are considered.}
\end{figure*}

\textit{Methodology---}To model hadron production in high-energy $\gamma Pb$ EMD at the LHC, we use the \texttt{Pythia} 8.316 Monte Carlo (MC) generator~\cite{Bierlich:2022pfr} with the \texttt{Angantyr} model~\cite{Bierlich:2018xfw} in the $\gamma Pb$ beam configuration.
This model describes the multiplicity and rapidity distributions of hadrons from $\gamma Pb$ interactions at the LHC reasonably well~\cite{Helenius:2024vdj}.
The photon can interact as a point-like
particle (direct photon) or it can fluctuate into a hadronic state (resolved photon); both of these subprocesses are included in \texttt{Pythia8}.
Figure~\ref{p8:dndeta}(a) shows the pseudorapidity distribution of produced charged particles in the simulated $\gamma Pb$ sample for various representative photon energies. While  particle production at low photon energies concentrates at relatively large pseudorapidity values, increasing the photon energy shifts the full distribution towards $\eta=0$. 
This demonstrates that the EMD process in UPCs with sufficiently high photon energy can produce particles (hadrons) even at mid-rapidity.

For a given set of exclusivity veto fiducial selection requirements, events generated using \texttt{Pythia8+Angantyr} are used to calculate veto-breaking probabilities ($p_{\text{vb}}$), defined as the probability of producing at least one hadron with certain requirements.
The $p_{\text{vb}}$ are parameterised as a function of incident photon energy ($E_{\gamma}$), calculated in the dissociated ion rest frame. This is demonstrated in Fig.~\ref{p8:dndeta}(b) for a charged particle veto with $p_T>100$~MeV and selected pseudorapidity requirements.
These veto-breaking probabilities are then used in selected UPC MC generators to correct the lowest-order EMD probability~\cite{Baltz:2009jk,Harland-Lang:2023ohq}:
\begin{align}
P^1_{Xn} (b) &=\int dE_{\gamma} \frac{d^3n_{\gamma}}{d^2bdE_{\gamma}} \sigma_{\gamma A }(E_{\gamma})\label{p1xn} & \\
\to P^{1,\text{v}}_{Xn} (b) &= \int dE_{\gamma} \frac{d^3n_{\gamma}}{d^2bdE_{\gamma}} \sigma_{\gamma A }(E_{\gamma}) (1-p_{\text{vb}}(E_{\gamma}))~,&
\end{align}
where $d^3n_{\gamma}/d^2bdE_{\gamma}$ is the photon emission flux from the ion and $\sigma_{\gamma A }(E_{\gamma})$ is the EMD cross section.
More precisely, $P^{1,\text{v}}_{Xn}$  is the mean number of excitations accompanied by no emission that would violate the exclusivity veto.
As the probability of having exactly $n$ EMD interactions follows a Poisson distribution, the probability of at least one
EMD interaction with no veto breaking is
\begin{equation}
    P^{\text{v}}_{Xn} (b) = (1-\exp{(-P^{1,\text{v}}_{Xn})})\exp{(-P^{1,\text{vb}}_{Xn})}~,
\end{equation}
where $P^{1,\text{vb}}_{Xn}=P^{1}_{Xn}-P^{1,\text{v}}_{Xn}$ is the mean number of EMD interactions that lead to exclusivity veto breaking and so the additional factor $\exp{(-P^{1,\text{vb}}_{Xn})}$ ensures there are no such interactions present.
We note that the probability of no EMD interactions,
\begin{equation}
    P^{\text{v}}_{0n} (b) = \exp{(-P^{1,\text{v}}_{Xn})}\exp{(-P^{1,\text{vb}}_{Xn})}=\exp{(-P^{1}_{Xn})}~,
\end{equation}
is not affected by the presence of any exclusivity vetos, as no particle production occurs here.

To evaluate $\sigma_{\gamma Pb }$, an approach typically used in MC generators relies on low-energy experimental data (available up to $E_{\gamma}=16$~GeV for $Pb$)~\cite{Klein:2016yzr, Harland-Lang:2018iur, Shao:2022cly}.
For $E_{\gamma}>16$~GeV, a Regge-based parameterization of the total photo-absorption cross sections of the proton at
high energy is used~\cite{ZEUS:2001wan}, scaled by the nuclear mass number times a nuclear suppression factor ($S_g$).
We use $S_g=0.4$ for our baseline results, as a similar suppression factor is reported in the measurements of light vector mesons in $PbPb$ UPC at the LHC~\cite{CMS:2025lsm, LHCb:2025fzk}. 

To estimate the model uncertainty, we follow the variations introduced in Ref.~\cite{Helenius:2024vdj} and vary the $p_{T,0}^{\mathrm{ref}}$ scale (which controls the multi-parton interaction probability in the case of resolved photons) between 3 and 4~GeV. We also explore alternative \texttt{Angantyr} subcollision models and photoproduction with a proton target. 
In addition, we vary the $\gamma Pb $ cross sections used in the MC models for $E_{\gamma} > 16$~GeV by $\pm50\%$. 
This variation is designed to cover the typical nuclear shadowing correction of $S_g\approx 0.6$ observed in the gluon nuclear parton distribution functions at small Bjorken-$x$~\cite{Duwentaster:2022kpv}.
Among these variations, as we will see in the following sections, by far the dominant effect is found to be due to the $S_g$ variation.
We note that these model uncertainties can be constrained by future measurements of rapidity gap cross sections in $\gamma Pb$ collisions at the LHC, in analogy to diffractive measurements performed in proton--proton collisions~\cite{ATLAS:2012djz}.

We will in addition in places include the impact of coincident $\rho^0$ production on the exclusivity veto. This process was recently observed for the first time by ATLAS~\cite{ATLAS:2025nac}, with the subsequent theoretical calculation of~\cite{Harland-Lang:2025bkk} providing a rather encouraging level of agreement to the observations. This is calculated following the approach of the latter paper, with the $\rho^0$ rapidity integrated over the appropriate interval and for simplicity 100\% acceptance assumed for the pionic decay, which is a good approximation.

We emphasize that in all existing MC implementations of UPCs, the impact of  EMD on the exclusivity veto is not accounted for. In terms of the EMD event fractions, 
in \texttt{SuperChic}~\cite{Harland-Lang:2023ohq}, a cut of 500 GeV is imposed on the photon energy (Eq.~\ref{p1xn}) to approximately match the requirement that the ion dissociation system lies in the $|y| > 5$ region. Although detailed implementations differ, other available MC models typically integrate the photon energy up to the kinematic limit~\cite{Broz:2019kpl,Shao:2022cly,Klein:2016yzr}, i.e. imposing no constraint at all.

Finally, we note that  a publicly available implementation of the veto described above is provided in the \texttt{SuperChic} MC, available at
\begin{center}
    \href{https://github.com/LucianHL/SuperChic}{https://github.com/LucianHL/SuperChic}.
\end{center}



\textit{Results: Exclusivity veto and ATLAS dimuon data---}We begin by considering the predicted impact of the exclusivity veto imposed by ATLAS in high precision measurements of exclusive dimuon production in $PbPb$ collisions at $\sqrt{s_{NN}}=5.02$ TeV~\cite{ATLAS:2020epq}. As discussed in detail in~\cite{Harland-Lang:2021ysd} the measured cross sections are found to lie on average roughly 10--15\% below the theoretical expectations, in particular as predicted by the \texttt{SuperChic} MC generator. The exclusivity veto effects discussed in the previous section provide a clear candidate to explain this discrepancy; namely, as the probability of the predicted signal being vetoed due to ion EMD is not accounted for then we may expect the corresponding predictions to be too high. 

\begin{figure}
\includegraphics[scale=0.6]{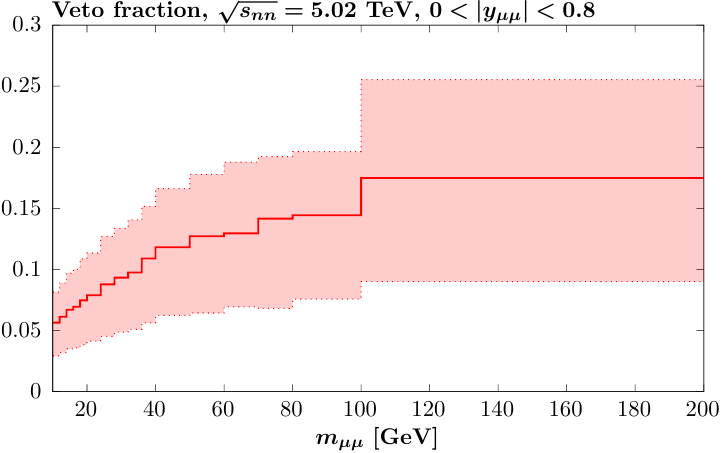}
\put(-120,-10){\textbf{(a)}}
\vspace{0.1in}
\hspace{0.3in}
\includegraphics[scale=0.6]{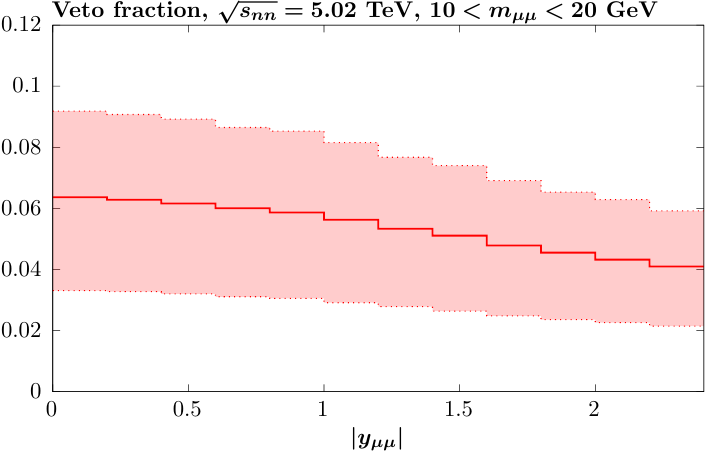}
\put(-120,-10){\textbf{(b)}}
\caption{\label{fig:ATLASdim_vetoefff} Predicted event fractions that fail the exclusivity veto imposed in the ATLAS measurement~\cite{ATLAS:2020epq} of exclusive dimuon production in $PbPb$ collisions at $\sqrt{s_{NN}}=5.02$ TeV, as a function of (a) dimuon invariant mass and (b) dimuon absolute rapidity, for a representative choice of dimuon system selection. The uncertainty band corresponds to varying the nuclear shadowing correction, as described in the text.}
\end{figure}

To quantify the expected impact of this effect, in Fig.~\ref{fig:ATLASdim_vetoefff} we show the predicted fraction of events that will fail the ATLAS exclusivity veto as a function of the dimuon rapidity and/or invariant mass, for a representative selection of the experimental regions. This is calculated following the approach described in the previous section, as implemented in \texttt{SuperChic}. In particular, the veto is accounted for following the $|\eta|<3.86$ curve in Fig.~\ref{p8:dndeta}(b) which reflects the acceptance of forward counters used in the ATLAS measurement~\cite{Sidoti:2014kra}. We can see that the predicted effect is noticeable, being at the 5--10\% level in the lower mass, central rapidity region, but being noticeably greater in other kinematic regions, approaching $\sim$10--20\% at high mass.
The significant kinematic dependence of this effect is particularly noteworthy. It arises from the impact parameter dependence of the EMD probability—as discussed in, e.g.,~\cite{Harland-Lang:2023ohq}—which leads to greater EMD at larger system invariant masses and more central rapidities.
This dependence clearly indicates that this cannot be straightforwardly corrected for in a manner that does not reference the underlying production process and its kinematic dependence. 

\begin{figure*}
\includegraphics[scale=0.6]{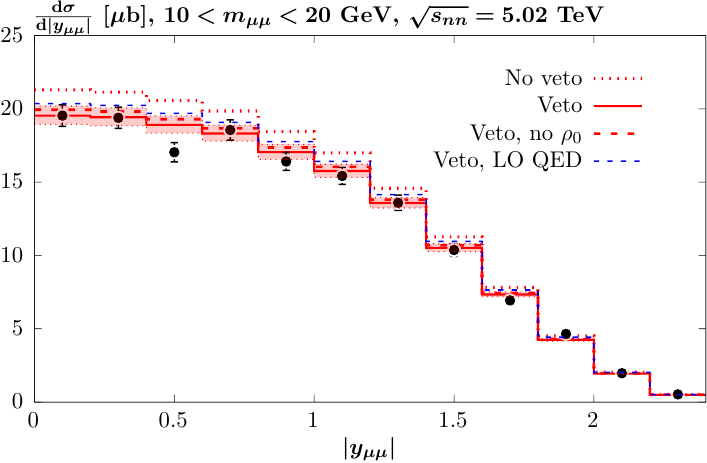}
\put(-120,-10){\textbf{(a)}}
\hspace{0.3in}
\includegraphics[scale=0.6]{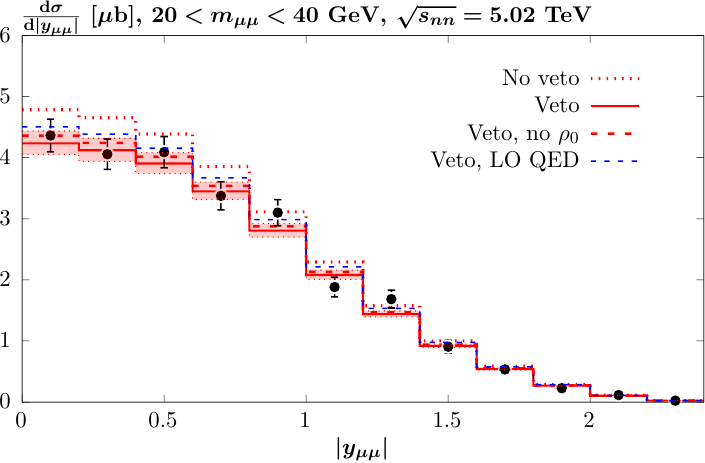}
\put(-120,-10){\textbf{(b)}} \\
\includegraphics[scale=0.6]{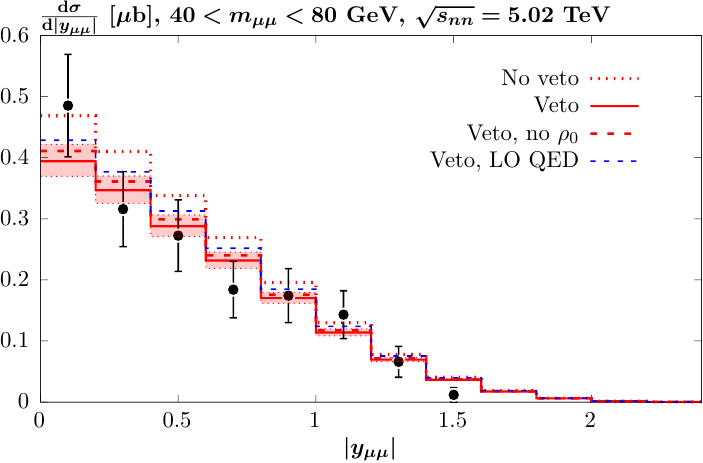}
\put(-120,-10){\textbf{(c)}}
\caption{\label{fig:ATLASdim_mvsdat} Predicted rapidity-differential cross sections compared to the  ATLAS measurement~\cite{ATLAS:2020epq} of exclusive dimuon production in $PbPb$ collisions at $\sqrt{s_{NN}}=5.02$ TeV, shown in different dimuon invariant mass regions: (a) $10<m_{\mu\mu}<20$~GeV, (b) $20<m_{\mu\mu}<40$~GeV and (c) $40<m_{\mu\mu}<80$~GeV, before and after accounting for the appropriate exclusivity veto for EMD hadrons. The uncertainty band corresponds to varying the nuclear shadowing correction, as described in the text. Results are calculated using \texttt{SuperChic+Pythia8+Angantyr} with NLO QED $k$-factors taken from~\cite{Shao:2024dmk}, and are also shown excluding $\rho^0$ production from the veto, and with a LO QED calculation for the $\gamma\gamma\to \mu\mu$ process are also shown, for the baseline value of the nuclear shadowing factor.}
\end{figure*}

The uncertainty band visible in Fig.~\ref{fig:ATLASdim_vetoefff} corresponds to varying the nuclear shadowing factor, $S_g$, by $\pm 50\%$, as described in the previous section. The impact of this variation is non-negligible, as we would expect given the veto fraction will scale roughly linearly with $S_g$, and the uncertainty on this is rather large. The impact of taking an alternative \texttt{Angantyr} subcollision model of photoproduction with a proton target is found to be significantly lower, at the percent level, and so is not included here. The impact of varying the $p_{T,0}^{\mathrm{ref}}$ scale is found to be  negligible.
Since the ATLAS event selection permits up to one forward counter hit, we systematically varied the veto selection to allow a maximum of one charged particle within the veto acceptance. 
This results in a relative shift of $-10\%$ in the correction, which we include as an additional model uncertainty.

In Fig.~\ref{fig:ATLASdim_mvsdat} we compare directly to the ATLAS data, before and after including the impact of the exclusivity veto on the predicted cross sections. We include NLO QED corrections as quoted in~\cite{Shao:2024dmk}. 
In particular, based on Fig.~7 of that reference, we assume negative corrections of 4\%, 6\%, and 8\% for the three dimuon invariant mass regions, in ascending order, and take these to be flat in dimuon rapidity.
We can see that including the EMD veto effects improves the description of the data markedly across all of the considered kinematic region. In the baseline result  we also include the effect of coincident $\rho^0$ production (not included in Fig.~\ref{fig:ATLASdim_vetoefff}), and indicate the impact of removing this for demonstration. This is found to be relatively minor, although not completely negligible. 

\begin{figure*}
\includegraphics[scale=0.6]{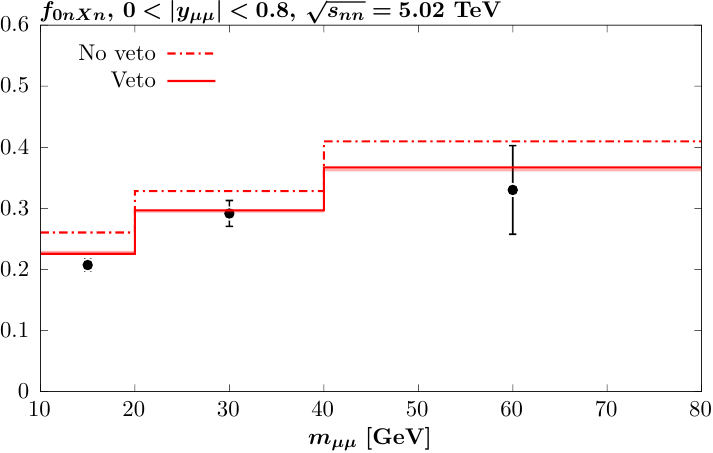}
\put(-120,-10){\textbf{(a)}}
\hspace{0.3in}
\includegraphics[scale=0.6]{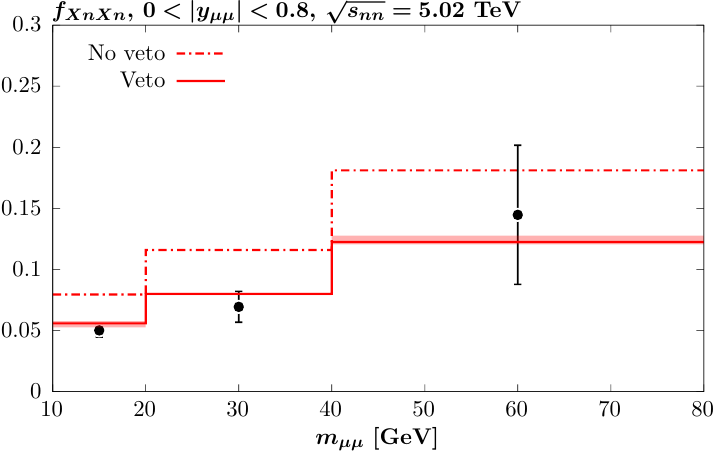}
\put(-120,-10){\textbf{(b)}}
\caption{\label{fig:f0X_ATLASdim_vsm} Impact of EMD hadron veto on (a) $0nXn$ and (b) $XnXn$ EMD fractions in the exclusive dimuon production for $|y_{\mu\mu}|<0.8$, as a function of $m_{\mu\mu}$. ATLAS data (points) are compared with \texttt{SuperChic+Pythia8+Angantyr} predictions with and without veto effects (solid and dashed lines, respectively). The uncertainty band on predictions corresponds to varying the nuclear shadowing correction, as described in the text.}
\end{figure*}

We next consider the impact of the exclusivity veto on the EMD event fractions that were also measured by ATLAS for the exclusive $\gamma\gamma\to\mu\mu$ process. In particular, as well as impacting on inclusive (w.r.t. ion EMD) production, the exclusivity veto will also affect the fraction of events with and without EMD, and hence these event fractions. In broad terms, the veto will effectively reduce the size of the EMD cross section and therefore suppress the predicted $0nXn$ and $XnXn$ event fractions. The results of including the exclusivity veto on the $0nXn$ and $XnXn$ EMD fractions are shown in Fig.~\ref{fig:f0X_ATLASdim_vsm} for the selected dimuon rapidity range. We can see again that including the impact of the veto significantly improves the description of the data such that a rather good description is achieved, up to some discrepancy in the lowest mass bin. We note that the predicted fractions in~\cite{Harland-Lang:2023ohq} were indeed observed to overshoot the data in a manner that qualitatively matches the `No veto' result shown here, but quantitatively are observed to lie somewhere between this and the `Veto' result; the reason for this is that in previous versions of \texttt{SuperChic}, the $E_\gamma > 500$ GeV photons were removed in Eq.~\ref{p1xn}, effectively accounting for some, but not all, of the effect of the exclusivity veto. 
The impact of varying the nuclear shadowing factor, $S_g$, is again shown; however, it is found to be rather milder than in the case of the veto fraction considered in Figs.~\ref{fig:ATLASdim_vetoefff}--\ref{fig:ATLASdim_mvsdat}, as expected for these ratio observables.


\textit{Results: Impact on the UPC coherent $J/\psi$ data---}Both the CMS and ALICE experiments have studied the dependence of the $\sigma(\gamma Pb\to J/\psi Pb)$ cross section on the photon--nucleus center-of-mass energy per nucleon ($W_{\gamma N}$)~\cite{CMS:2023snh, ALICE:2023jgu}. The extraction of $\sigma(\gamma Pb\to J/\psi Pb)$ 
is performed by a simultaneous fit to the rapidity-dependent exclusive $J/\psi$ cross sections across different EMD classes ($0n0n$, $0nXn$ and $XnXn$). These classes provide distinct photon flux configurations, allowing the fit to decouple the low-energy and high-energy photon contributions that otherwise overlap in the summed rapidity distributions.
However, as demonstrated in the previous section, hadrons from high-energy EMD can severely impact the exclusivity veto in the $0nXn$ and $XnXn$ events, which is not (at least fully) accounted for in these studies.
To check the impact of the exclusivity veto  on the $0nXn$ and $XnXn$ photon fluxes (labeled as $n^{\gamma}_{0nXn}$ and $n^{\gamma}_{XnXn}$ respectively), we re-calculate $n^{\gamma}_{0nXn}$ and $n^{\gamma}_{XnXn}$ for UPC $J/\psi$ production using a modified version of the \texttt{STARlight} MC generator~\cite{Klein:2016yzr}, taking into account experimental fiducial veto requirements. Similarly to the \texttt{SuperChic} based studies in the previous section, we use parameterised veto-breaking probabilities from the \texttt{Pythia8+Angantyr} $\gamma Pb$ model.

For the CMS measurement we follow the  fiducial veto requirements: no extra charged particles for $|\eta|<2.4$ and $p_T\gtrsim 500$~MeV and no extra particles with $E>7.5$~GeV and $3<|\eta|<5.2$~\cite{CMS:2023snh}.
For the ALICE data, due to the asymmetric $\eta$ requirements of counters on both detector sides, we use the averaged value of $|\eta|<6.3$ from counters with the largest value of $|\eta|$~\cite{Broz:2020ejr}.
Here we also assume a $p_T$ threshold for charged particles of 100~MeV; we point out, however, that the momentum acceptance of ALICE counters is not precisely known.
The extracted $p_{\text{vb}}(E_{\gamma})$ curves for the CMS and ALICE veto definitions are shifted toward lower energies relative to the $|\eta|<5$ curve in Fig.~\ref{p8:dndeta}(b), representing a more restrictive veto at lower $E_{\gamma}$. 

To check the impact of EMD hadrons on $\sigma(\gamma Pb\to J/\psi Pb)$, CMS and ALICE data are refitted with the improved values of $n^{\gamma}_{0nXn}$ and $n^{\gamma}_{XnXn}$. According to \texttt{STARlight+Pythia8+Angantyr} predictions, the values of $n^{\gamma}_{0nXn}$ are reduced by about 20--25\%, and $n^{\gamma}_{XnXn}$ by about 45--50\% (with respect to default \texttt{STARlight} predictions) due to the EMD veto-breaking effect. It should be noted that these variations are not covered by the assigned photon flux uncertainties in the original CMS and ALICE fits.

We point out that the ALICE measurement partly mitigates this effect by correcting the measured $0nXn$ and $XnXn$ $J/\psi$ cross sections using correction factors extracted from inclusive EMD events.
In particular, ALICE measures the following correction factors in the forward ($2.5<|y_{J/\psi}|<4$) $J/\psi$ measurement: $\epsilon^{0nXn}_{\mathrm{EMD}}=0.88$ and $\epsilon^{XnXn}_{\mathrm{EMD}}=0.84$~\cite{ALICE:2023jgu}.
Comparing these numbers with the flux suppression numbers above, we find that the former are significantly underestimated, as much smaller impact parameters are probed in UPC $J/\psi$+EMD events compared to the inclusive EMD process.
Consequently, we have replaced the original ALICE EMD veto corrections here by the predictions derived from the \texttt{STARlight+Pythia8+Angantyr} model.

\begin{figure}
\includegraphics[scale=0.45]{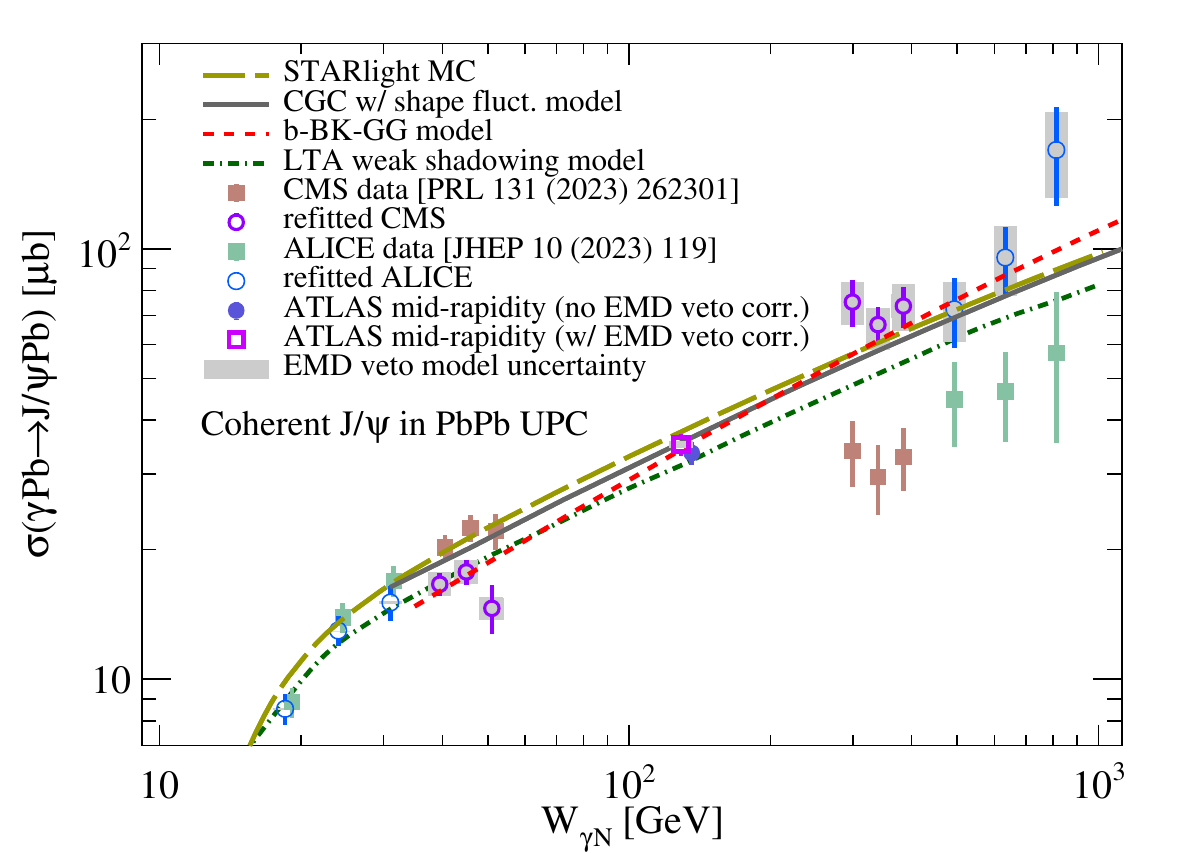}
\caption{\label{cms:jpsi} The total coherent UPC $J/\psi$ photoproduction cross section as a function of $W_{\gamma N}$ from the original and refitted CMS and ALICE forward-$J/\psi$ data, and ATLAS mid-rapidity data. Data are presented together with the predictions from \texttt{STARlight} MC generator~\cite{Klein:2016yzr}, color glass condensate (CGC) model~\cite{Mantysaari:2023xcu}, the b-BK-GG gluon saturation model~\cite{Bendova:2020hbb}, and the leading twist approximation (LTA) weak shadowing model~\cite{Guzey:2024gff}. Vertical bars denote the original set of CMS/ALICE/ATLAS data uncertainties, while the grey boxes represent the EMD veto model uncertainty. }
\end{figure}

Figure~\ref{cms:jpsi} presents the comparison of the original CMS and ALICE fits and our new fits, with corrected photon fluxes. 
The EMD veto model uncertainty (shown as gray boxes) is again dominated by the $S_g$ uncertainty. Due to the strong sensitivity of high-$W_{\gamma N}$ data on $n^{\gamma}_{XnXn}$, the values of $\sigma(\gamma Pb\to J/\psi Pb)$ for high-$W_{\gamma N}$ increase by about a factor of two, as the $n^{\gamma}_{XnXn}$ fluxes decrease by a similar amount after applying the EMD veto corrections.
We only consider the forward ($2.5<|y_{J/\psi}|<4$) selection of the ALICE measurement, as the analogous measurement at mid-rapidity ($|y_{J/\psi}|<0.8$) may be significantly affected by the particle veto bias effect due to low-energy electrons from the accompanying $\gamma\gamma\to ee$ process (as suggested in Ref.~\cite{ATLAS:2025aav}), which we do not simulate within the current setup.
In addition, we include ATLAS mid-rapidity data~\cite{ATLAS:2025aav}, with and without EMD veto corrections, extracted using the relation: $\sigma(\gamma Pb\to J/\psi Pb)=\frac{1}{2n^{\gamma}}\frac{d\sigma_{J/\psi}}{dy}|_{y=0}$, where $n^{\gamma}$ is the EMD-inclusive photon flux from \texttt{STARlight} and $d\sigma_{J/\psi}/{dy}|_{y=0}$ is the measured exclusive coherent $J/\psi$ cross section extrapolated to $y=0$.

In Fig.~\ref{cms:jpsi} data are compared with selected models describing the $\sigma(\gamma Pb\to J/\psi Pb)$ evolution~\cite{Klein:2016yzr,Mantysaari:2023xcu,Bendova:2020hbb, Guzey:2024gff}.
As can be seen, the procedure largely resolves the tension between the models and the LHC data, particularly in the high-$W_{\gamma N}$ region, corresponding to small Bjorken-$x$, where parton saturation effects may become relevant.
In particular, the b-BK-GG gluon saturation model~\cite{Bendova:2020hbb} describes the corrected data reasonably well.

\textit{Conclusions---}In summary, we provide a methodology to calculate corrections for ultraperipheral interaction processes with exclusive final states at the LHC, accounting for hadron production from accompanying electromagnetic dissociation (EMD). 
These corrections affect all such LHC measurements with exclusive final states and can significantly suppress the relevant cross sections. These are found to depend sensitively on the production process and kinematics, implying that a precise account of the exclusivity veto in theoretical predictions is mandatory. Once applied, they resolve apparent data-to-theory 
tensions for  high-mass $\gamma\gamma \to \mu\mu$ and coherent $\gamma Pb \to J/\psi Pb$ UPC processes.
As the magnitude of these corrections depends on the $\gamma Pb$ cross sections with high-energy photons, a detailed LHC measurement of this process could help to better constrain them.

\textit{Acknowledgments---}MD thanks the National Science Center of Poland for support via the grant award UMO-2022/47/O/ST2/00148. LHL thanks the Science and Technology Facilities Council (STFC) part of U.K.
Research and Innovation for support via the grant award ST/T000856/1. 

\onecolumngrid
\appendix
\section*{Supplemental Material}

\subsection{$\gamma\gamma\to\mu\mu$ breakup fractions for other dimuon rapidity ranges}
\begin{figure}[b]
\includegraphics[scale=0.6]{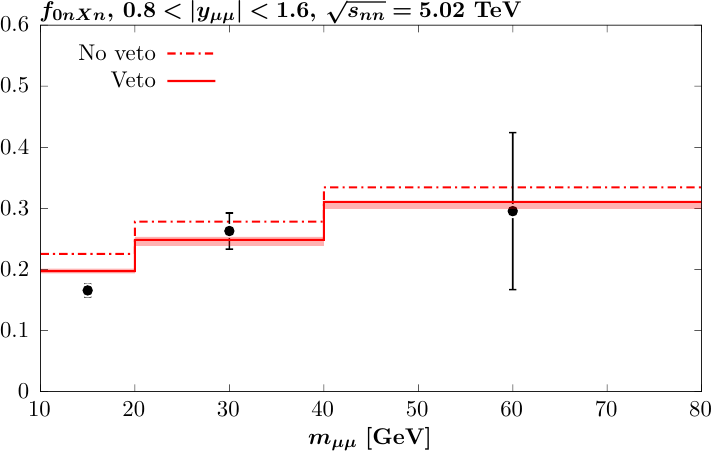}
\put(-120,-10){\textbf{(a)}} 
\hspace{0.3in}
\includegraphics[scale=0.6]{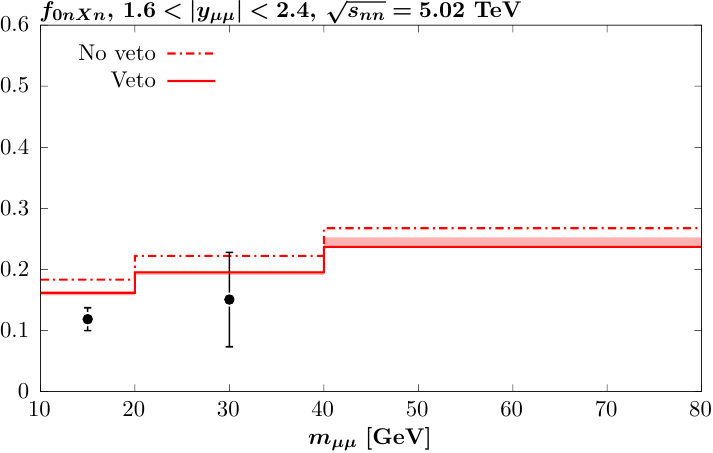}
\put(-120,-10){\textbf{(b)}}
\caption{\label{fig:f0X_ATLASdim_vsm_other_y} Impact of EMD hadron veto on $0nXn$ EMD fractions in the exclusive dimuon production, as a function of $m_{\mu\mu}$, in other dimuon rapidity regions: (a) $0.8<|y_{\mu\mu}|<1.6$ and (b) $1.6<|y_{\mu\mu}|<2.4$. ATLAS data (points) are compared with \texttt{SuperChic+Pythia8+Angantyr} predictions with and without veto effects (solid and dashed lines, respectively). The uncertainty band on predictions corresponds to varying the nuclear shadowing correction, as described in the text.}
\end{figure}

\begin{figure}[]
\includegraphics[scale=0.6]{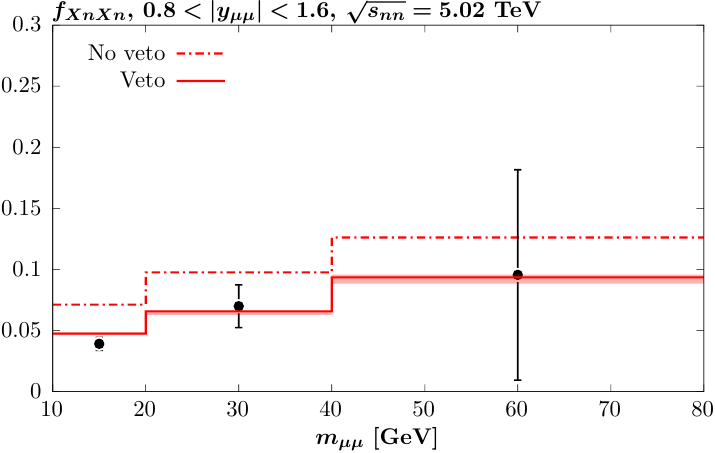}
\put(-120,-10){\textbf{(a)}} 
\hspace{0.3in}
\includegraphics[scale=0.6]{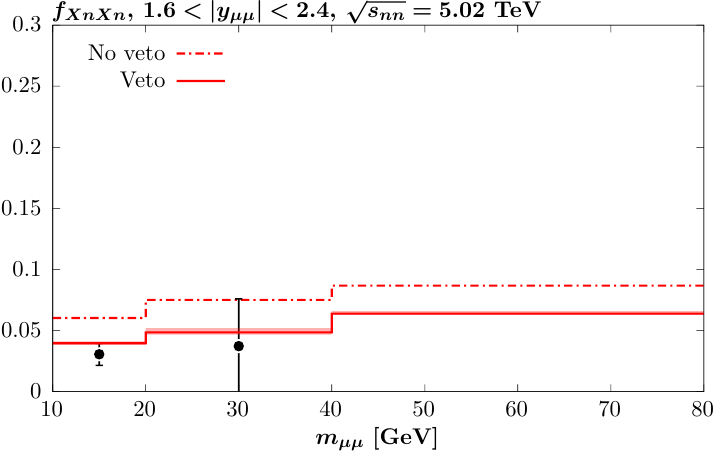}
\put(-120,-10){\textbf{(b)}}
\caption{\label{fig:fXX_ATLASdim_vsm_other_y}  Impact of EMD hadron veto on $XnXn$ EMD fractions in the exclusive dimuon production, as a function of $m_{\mu\mu}$, in other dimuon rapidity regions: (a) $0.8<|y_{\mu\mu}|<1.6$ and (b) $1.6<|y_{\mu\mu}|<2.4$. ATLAS data (points) are compared with \texttt{SuperChic+Pythia8+Angantyr} predictions with and without veto effects (solid and dashed lines, respectively). The uncertainty band on predictions corresponds to varying the nuclear shadowing correction, as described in the text.}
\end{figure}

Figures~\ref{fig:f0X_ATLASdim_vsm_other_y} and \ref{fig:fXX_ATLASdim_vsm_other_y} show the data-to-model comparisons for $0nXn$ and $XnXn$ EMD event fractions for the exclusive dimuon production, as a function of $m_{\mu\mu}$, in other dimuon rapidity regions: $0.8<|y_{\mu\mu}|<1.6$ and $1.6<|y_{\mu\mu}|<2.4$.

\subsection{Study of the nuclear suppression factor with UPC coherent $J/\psi$}

It is often convenient to convert the $\sigma(\gamma Pb\to J/\psi Pb)$ photoproduction cross sections into
the nuclear suppression factor $S_g$, which can be defined as:
\begin{equation}
    S_g (x) = \sqrt{\frac{\sigma(\gamma Pb\to J/\psi Pb)}{\sigma_{IA}(\gamma Pb\to J/\psi Pb)}}~,
\end{equation}
where $x=m_{J/\psi}^2/W_{\gamma N}^2$ and  $\sigma_{IA}(\gamma Pb\to J/\psi Pb)$ term denotes the cross sections from the impulse approximation (IA) model~\cite{Guzey:2013xba}. This model is based on a simple scaling of experimental data from exclusive $J/\psi$ photoproduction off protons and neglects the nuclear modifications of the gluon density.
One can identify $S_g(x)$ with $R_g(x,Q^2_0) = g_A(x,Q^2_
0)/[Ag_p(x,Q^2_0)]$ representing the ratio of the nuclear to the proton gluon density, where $Q^2_0= 3$ GeV$^2$ is the resolution scale set by the charm quark mass, which allows to interpret the UPC $J/\psi$ data in terms of nuclear PDFs~\cite{Guzey:2013xba,Guzey:2013qza,Guzey:2024gff}.

\begin{figure}[h!]
\includegraphics[scale=0.45]{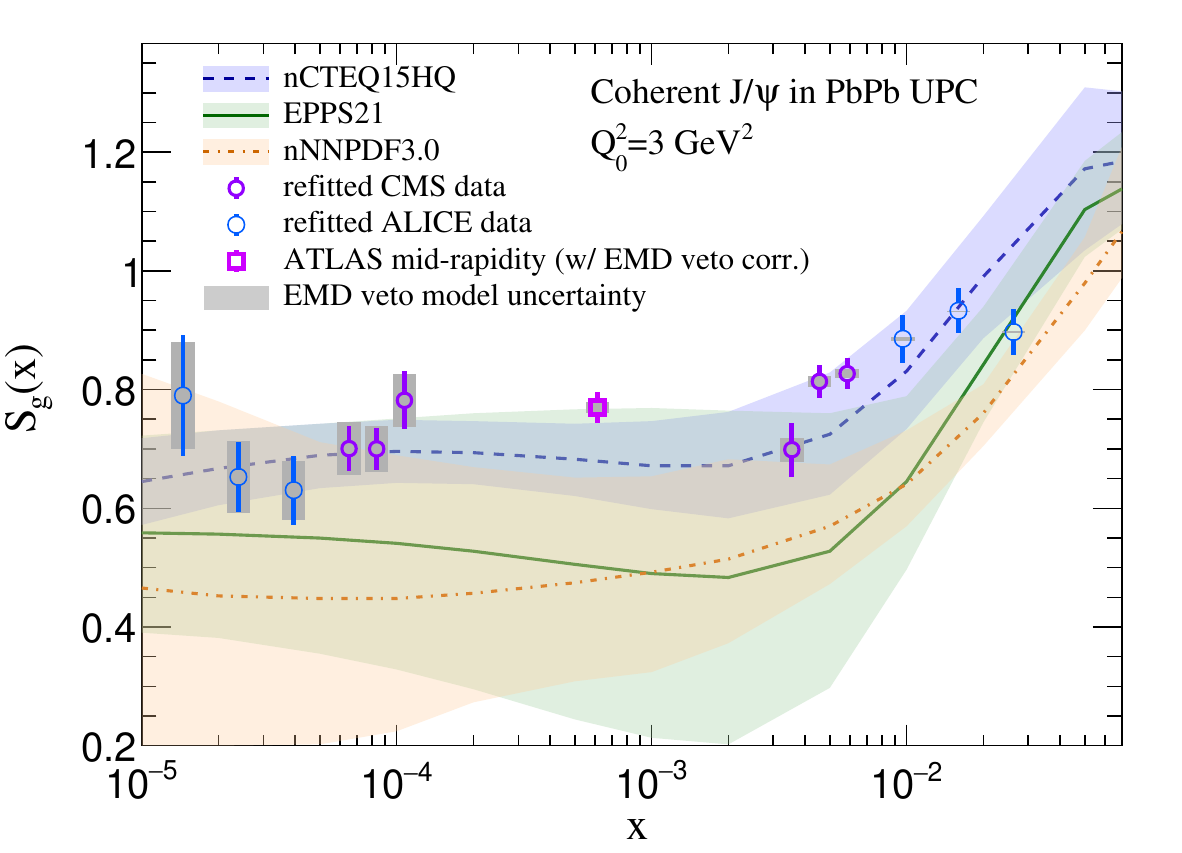}
\caption{\label{cms:jpsisupp} 
The nuclear suppression factor $S_g$ as a function of $x$ for the EMD-veto-corrected CMS, ALICE and ATLAS data on coherent exclusive $J/\psi$ production in $PbPb$, in comparison with theoretical predictions using the recent nuclear PDF parameterisations (EPPS21~\cite{Eskola:2021nhw}, nCTEQ15HQ~\cite{Duwentaster:2022kpv}, nNNPDF3.0~\cite{AbdulKhalek:2022fyi}) calculated for $Pb$ at $Q^2_0= 3$ GeV$^2$. Vertical bars denote the original set of CMS/ALICE/ATLAS data uncertainties, while the grey boxes represent the EMD veto model uncertainty. 
Coloured shaded regions represent nuclear PDF uncertainties.}
\end{figure}

Figure~\ref{cms:jpsisupp} presents the $S_g$	
  values extracted from the EMD-veto-corrected CMS, ALICE, and ATLAS coherent exclusive $J/\psi$ data shown in Fig.~\ref{cms:jpsi}, as a function of $x$. The corrected data are compared with theoretical predictions for $g_A/(Ag_p)$ for $Pb$ at $Q^2_0= 3$ GeV$^2$, obtained using the EPPS21~\cite{Eskola:2021nhw}, nCTEQ15HQ~\cite{Duwentaster:2022kpv}, and nNNPDF3.0~\cite{AbdulKhalek:2022fyi} nuclear PDFs.
It can be seen that, within theoretical uncertainties, these modern nuclear PDF predictions agree with the corrected data, exhibiting essentially flat behavior for $S_g(x)$ at $x<0.005$. We note that this trend is not observed in the original data (prior to the EMD veto correction), as discussed in Ref.~\cite{Guzey:2024gff}.

\twocolumngrid
\bibliography{apssamp}

\end{document}